\documentstyle[aps,prl,epsf,floats,amsfonts]{revtex}
\newcommand{\beq}{\begin{equation}}
\newcommand{\eeq}{\end{equation}}
\newcommand{\beqs}{\begin{eqnarray}}
\newcommand{\eeqs}{\end{eqnarray}}

\newcommand{\lsim}{\mathrel{\raisebox{-.6ex}{$\stackrel{\textstyle<}{\sim}$}}}
\newcommand{\gsim}{\mathrel{\raisebox{-.6ex}{$\stackrel{\textstyle>}{\sim}$}}}

\draft

\tighten
 
\begin{document}

\twocolumn[\hsize\textwidth\columnwidth\hsize\csname
@twocolumnfalse\endcsname

\title{$n - \bar n$ Oscillations in Models with Large Extra Dimensions}

\author{Shmuel Nussinov$^1$ and Robert Shrock$^2$} \address{$^1$ \ Sackler
Faculty of Science, Tel Aviv University, Tel Aviv, Israel \\ 
$^2$ \ C. N. Yang Institute for Theoretical Physics, State University of New 
York, Stony Brook, NY 11794}

\baselineskip 6.0mm
 
\tighten

\maketitle
 
\begin{abstract}

We analyze $n - \bar n$ oscillations in generic models with large extra
dimensions in which standard-model fields propagate and fermion wavefunctions
have strong localization.  We find that in these models $n - \bar n$ 
oscillations might occur at levels not too far below the current limit. 

\end{abstract}


\vskip2.0pc]

Although current experimental data are fully consistent with a four-dimensional
Minkowski spacetime, it is useful to explore the possibility of extra
dimensions, both from a purely phenomenological point of view and because the
main candidate theory for quantum gravity - string theory - suggests the
existence of higher dimensions.  Here we shall focus on theories in which the
standard-model (SM) fields can propagate in the extra dimensions and the
wavefunctions of the SM fermions have strong localization (with Gaussian
profiles) at various points in this extra-dimensional space
\cite{as}-~\cite{bbgr}.  Such models are of interest partly because they can
provide a mechanism for obtaining a hierarchy in fermion masses and quark
mixing.  In generic models of this type, excessively rapid proton decay can be
avoided by arranging that the wavefunction centers of the $u$ and $d$ quarks
are separated far from those of the $e$ and $\mu$ \cite{as}.  However, this
separation does not, by itself, guarantee adequate suppression of another
source of baryon number violation, namely $n - \bar n$ oscillations.  Here we
shall analyze these oscillations in generic models of this type.  Early studies
of $n-\bar n$ oscillations in conventional $d=4$ dimensional spacetime include
\cite{kuzmin}-\cite{rs2}; there is currently renewed experimental and
theoretical interest \cite{ornl,bm}.

Our theoretical framework is as follows. Usual spacetime coordinates are
denoted as $x_\nu$, $\nu=0,1,2,3$ and the $n$ extra coordinates as $y_\lambda$;
for definiteness, the latter are taken to be compact.  The framework is such
that fermion fields have the form $\Psi(x,y)=\psi(x)\chi(y)$.  In the extra
dimensions the SM fields are assumed to have support in an interval $0 \le
y_\lambda \le L$.  We define $\Lambda_L \equiv L^{-1}$.  The
$d=(4+\ell)$-dimensional fields thus have Kaluza-Klein (KK) mode
decompositions.  We shall work in a low-energy effective field theory approach
with an ultraviolet cutoff $M_*$.  These models provide a possible explanation
for the hierarchy in the fermion mass matrices via the localization of fermion
wavefunctions with half-width $\mu^{-1} << L$ at various points in the
higher-dimensional space.  We denote $\xi=\mu L$; $\xi \sim 30$ is chosen for
adequate separation of the various fermion wavefunctions while still fitting
well within the thick brane.  As a result of this localization, the
$y$-dependent part of the wavefunction for a fermion field $f$ has the generic
form $\chi_f(y) = Ae^{-\mu^2|y-y_f|^2}$, where $y_f$ denotes the position
vector of this fermion in the extra dimensions, $|y_f| = (\sum_{\lambda=1}^\ell
y_{f,\lambda}^2)^{1/2}$.  For $\ell=1$ and $\ell=2$, this fermion localization
can be accomplished in a low-energy field-theoretic manner by coupling to a
scalar with a kink or vortex solution, respectively \cite{trap}. The
normalization factor $A=(2/\pi)^{\ell/4}\mu^{\ell/2}$ is included so that after
the integration over the $\ell$ extra dimensions, the kinetic term $\bar
\psi(x) i \partial \llap/ \psi(x)$ has canonical normalization.  Starting from
an effective Lagrangian in the $d$-dimensional spacetime, one obtains the
resultant effective theory in four dimensions by integrating over the extra
$\ell$ dimensions.  In particular, starting from a Yukawa interaction in the
$d$-dimensional space with coefficients of order unity and integrating over the
$\ell$ extra coordinates, using the fact that the convolution of two of the
fermion Gaussian wavefunctions is another Gaussian, one finds that the
coefficients of the resultant 4-dimensional Yukawa interaction contain factors
$e^{-(\mu^2/2)|y_f-y_{f^\prime}|^2}$, so that moderate spatial separations
$|y_f-y_{f^\prime}|$ for localized fermion wavefunctions can produce strong
hierarchies in the 4D Yukawa matrices.  We focus here on the case $d=6$, i.e.,
$\ell=2$, for which a fit to the quark masses and CKM matrix was obtained
\cite{as,bgr}.  The UV cutoff $M_*$ satisfies $M_* > \mu$ for the validity the
low-energy field theory analysis and $M_*/\Lambda_L \sim 10^3$ for the
perturbativity of the top quark Yukawa coupling.  Further, $\Lambda_L \gsim
100$ TeV for adequate suppression of neutral flavor-changing current (NFCC)
processes \cite{dpq,c1,c2}; with the ratio $\xi=30$, this yields $\mu \sim 3
\times 10^3$ TeV.

In field-free vacuum, the $2 \times 2$ Hamiltonian ${\cal H}$ in the $(n, \bar
n)$ space has matrix elements $\langle n | {\cal H} | n \rangle = 
\langle \bar n | {\cal H} | \bar n \rangle = m_n$ (assuming CPT conservation)
and $\langle \bar n | {\cal H} | n \rangle = 
    \langle \bar n | {\cal H}_{eff} | n \rangle \equiv \delta m$, where 
${\cal H}_{eff}$ denotes the interaction responsible for the $n- \bar n$
oscillations.  This leads to the oscillation probability $|\langle \bar n
| n(t) \rangle|^2 = \sin^2 (t/\tau_{n \bar n})$, where $\tau_{n \bar n} =
1/|\delta m|$, and it is straightforward to generalize this to the case where
the earth's magnetic field is included. Searches with free neutrons from
reactors have yielded the lower limit $\tau_{n \bar n} \ge 0.86 \times 10^8$
sec \cite{grenoble}.  Further, the $n - \bar n$ oscillations cause matter
instability since the $\bar n$ annihilates with surrounding matter and, related
to this via crossing, two neutrons can annihilate to pions.  Although the $n -
\bar n$ mixing in matter is strongly suppressed by the fact that the diagonal
matrix elements $\langle n | {\cal H} | n \rangle = m_n+V_n$ and $\langle
\bar n | {\cal H} | \bar n \rangle = m_n + Re(V_{\bar n}) + i Im(V_{\bar
n}) $ are different, this is compensated for by the large number of nucleons in
modern nucleon decay detectors.  Searches using these detectors have yielded
the limit for matter instability (m.i.) caused by this process (leading to
multipion final states) of $\tau_{m.i.} \gsim 0.6 \times 10^{32}$ yrs which,
with reasonable inputs for the nuclear potentials $V_n$ and $V_{\bar n}$, have
yielded a current limit very close to that from searches with free $n$'s:
$\tau_{n \bar n} \ge 1.2 \times 10^8$ sec, i.e., $|\delta m| \le 0.6 \times
10^{-32}$ GeV \cite{mat_inst}.

In 4D, ${\cal H}_{eff}$ consists of a sum of six-quark operators ${\cal O}_i$
with coefficients $c_i$ having (mass) dimension $-5$.  If, as is the case here,
the effective scale relevant for $n-\bar n$ oscillations is large compared with
the electroweak symmetry-breaking scale of $\sim 250$ GeV, then the ${\cal
O}_i$ must be invariant under $G_{SM} =$ SU(3) $\times$ SU(2)$_L \times $
U(1)$_Y$.  There are four relevant (linearly independent) ${\cal O}_i$ of this
type, so that ${\cal H}_{eff}=\sum_{i=1}^4 c_i {\cal O}_i$, where the ${\cal
O}_i$ can be chosen to be
\beq
{\cal O}_1 = [u_R^{\alpha T} C u_R^\beta][d_R^{\gamma T} C d_R^\delta]
[d_R^{\rho T} C d_R^\sigma ](T_s)_{\alpha \beta \gamma \delta \gamma \rho
\sigma} 
\label{p1}
\eeq
\beq
{\cal O}_2 = [u_R^{\alpha T} C d_R^\beta][u_R^{\gamma T} C d_R^\delta]
[d_R^{\rho T} C d_R^\sigma ](T_s)_{\alpha \beta \gamma \delta \gamma \rho
\sigma}
\label{p2}
\eeq
\beq
{\cal O}_3 = [Q_L^{i \alpha T} C Q_L^{j \beta}][u_R^{\gamma T} C d_R^\delta]
[d_R^{\rho T} C d_R^\sigma ]\epsilon_{ij} 
(T_a)_{\alpha \beta \gamma \delta \gamma \rho \sigma}
\label{p3}
\eeq
\beq
{\cal O}_4 = [Q_L^{i \alpha T} C Q_L^{j \beta}]
             [Q_L^{k \gamma T} C Q_L^{m \delta}]
             [d_R^{\rho T} C d_R^\sigma ]\epsilon_{ij}\epsilon_{km} 
(T_a)_{\alpha \beta \gamma \delta \gamma \rho \sigma}
\label{p5}
\eeq
where $Q_L = {u \choose d}_L$, Greek and Latin indices are color SU(3) and weak
SU(2) indices, respectively, $C$ is the Dirac charge conjugation matrix, and 
\beq
(T_a)_{\alpha \beta \gamma \delta \gamma \rho \sigma} = 
\epsilon_{\rho \alpha \beta}\epsilon_{\sigma \gamma \delta} +
\epsilon_{\sigma \alpha \beta}\epsilon_{\rho \gamma \delta}
\label{ta}
\eeq
\beq
(T_s)_{\alpha \beta \gamma \delta \gamma \rho \sigma} =
\epsilon_{\rho \alpha \gamma}\epsilon_{\sigma \beta \delta} +
\epsilon_{\sigma \alpha \gamma}\epsilon_{\rho \beta \delta} +
\epsilon_{\rho \beta \gamma}\epsilon_{\sigma \alpha \delta} +
\epsilon_{\sigma \beta \gamma}\epsilon_{\rho \alpha \delta}
\label{ts}
\eeq
The matrix elements of $\langle \bar n | {\cal O}_i | n \rangle$ for $1 \le i
\le 4$ were calculated in the MIT bag model in \cite{rs}. 

In the higher-dimensional field theory, we denote the effective Hamiltonian
responsible for the $n- \bar n$ transitions as ${\cal H}_{eff,4+n}=
\sum_{i=1}^4 \kappa_i O_i$, where the operators $O_i$ are comprised of the
$(4+\ell)$-dimensional quark fields corresponding to those in ${\cal O}_i$ as
$\Psi$ corresponds to $\psi$. The coefficient of an operator product consisting
of $N_\Psi$ fermions has (mass) dimension ${\rm dim}[c]=d-(N_\Psi/2)(d-1)$, so
that ${\rm dim}[c]=-9$ for $d=6$ and $N_\Psi=6$.  We denote the effective scale
of physics responsible for the $n -\bar n$ oscillations as $M_X$ and thus, for
the $d=6$ case of interest here, write $\kappa_i=\eta_i/M_X^9$, where the
$\eta_i$'s are dimensionless constants.  It will be convenient, and will incur
no loss of generality, to define $M_X$ so that $\eta_4 \equiv 1$.

We next carry out the integrations over the extra dimensions to get, for each
$i$, 
\beq
c_i{\cal O}_i(x) = \kappa_i \int d^2y O_i(x,y)
\label{oint}
\eeq
Letting $\rho_c \equiv 4\mu^4/(3\pi^2 M_X^9)$, we find 
\beq c_i = \rho_c \eta_i \exp \left [-(4/3)\mu^2|y_{u_R}-y_{d_R}|^2 \right ] \
, i=1,2
\label{c12}
\eeq
\beqs
c_3 &=& \rho_c \eta_3 \exp [ -(1/6)\mu^2(2|y_{Q_L}-y_{u_R}|^2
+6|y_{Q_L}-y_{d_R}|^2 \cr\cr
& + & 3|y_{u_R}-y_{d_R}|^2) ]
\label{c3}
\eeqs
\beq
c_4 = \rho_c \exp \left [-(4/3) \mu^2|y_{Q_L}-y_{d_R}|^2 \right ]
\label{c4}
\eeq
A fit to data for $n=2$ yielded $|y_{Q_L}-y_{u_R}| = |y_{Q_L}-y_{d_R}| \simeq
5\mu^{-1}$ and $|y_{u_R}-y_{d_R}| \simeq 7\mu^{-1}$ \cite{as}.  One can also
include corrections due to Coulombic gauge interactions between fermions
\cite{qlx}.  Using these inputs, we find that $c_j$ for $j=1,2,3$ are
negligibly small compared with $c_4$, and we hence focus on $c_4$.  Neglecting
small CKM mixings, the quantity $|y_{Q_L}-y_{d_R}|$ is determined by $m_d$ via
the relation $m_d = h_d (v/\sqrt{2})$ with $h_d = h_{d,0}\exp (-(1/2)\mu^2
|y_{Q_L}-y_{d_R}|^2)$, where $h_{d,0}$ is the Yukawa coupling in the
$(4+\ell)$-dimensional space, whence 
\beq
\exp \left [ -(1/2)\mu^2|y_{Q_L}-y_{d_R}|^2 \right ]  =
\frac{2^{1/2}m_d}{h_{d,0}v}
\label{mdrel}
\eeq
We take $h_{d,0} \sim 1$, and $m_d \simeq 10$ MeV.  Hence, the contribution to
$\delta m$ from the ${\cal O}_4$ term is
\beqs
\delta m & \simeq & c_4 \langle \bar n | {\cal O}_4 | n \rangle \cr\cr
& \simeq & \left ( \frac{4\mu^4}{3\pi^2 M_X^9} \right ) 
\left ( \frac{2^{1/2}m_d}{v} \right )^{8/3}
\langle \bar n | {\cal O}_4 | n \rangle
\label{deltam}
\eeqs
The matrix element $\langle \bar n | {\cal O}_4| n \rangle$ ($=4\langle {\cal
O}_3 \rangle_{LLR}$ in the notation of \cite{rs,rs2}) was calculated to be $1
\times 10^{-4}$ and $0.8 \times 10^{-4}$ GeV$^6$ in the MIT bag model for its
two main parameter fits \cite{rs}; here we average these.  Requiring that the
resultant value of $|\delta m|$ be less than the experimental limit,
$1/(1.2 \times 10^8 {\rm sec}) = 0.55 \times 10^{-32}$ GeV
\cite{grenoble,mat_inst}, we obtain the bound
\beqs
M_X & \gsim & (45 \ {\rm TeV}) 
\left ( \frac{\tau_{n \bar n}}{1.2 \times 10^8 \ {\rm sec} } \right )^{1/9} 
\cr\cr & \times & \left (\frac{\mu}{3 \times 10^3 \ {\rm TeV}} \right )^{4/9}
\left ( \frac{|\langle \bar n | O_4| n \rangle |}
{0.9 \times 10^{-4} \ {\rm GeV}^6} \right )^{1/9}
\label{lambdalow2}
\eeqs
The uncertainty in the calculation of the matrix element
$\langle \bar n | {\cal O}_4 | n \rangle$ is relatively unimportant for our
bound because of the $1/9$ power in (\ref{lambdalow2}) \cite{lgt}. 

The result (\ref{lambdalow2}) is very interesting, since it shows that for
values of $M_X$ in the range relevant to extra-dimensional models of the type
considered here, $n- \bar n$ oscillations might occur at levels that are in
accord with the current experiment limit but not too far below this limit. Our
finding provides motivation for further searches for $n - \bar n$ oscillations,
e.g. via an analysis of SuperKamiokande data or a new free-neutron reactor
experiment with increased sensitivity, such as that proposed in \cite{ornl}.
Considering only the analysis of existing data from SuperK, a rough estimate is
that the search for matter instability via $n-\bar n$ oscillations followed by
annihilation to multipion final states might be able to set a lower bound on
$\tau_{m.i.}$ in the range of $10^{33}$ yrs, similar to bounds that have been
set by SuperK on various proton decay modes; this could increase the lower
bound on $\tau_{n \bar n}$ to $\sim 3 \times 10^8$ sec.  If a signal were to be
seen, other experimental data and analysis would be necessary to determine
whether these oscillations are associated with extra dimensions, since, as has
been shown in \cite{bm}, they can also occur with $\tau_{n \bar n} \lsim 10^9 -
10^{10}$ sec in certain four-dimensional supersymmetric models.

S. N. thanks the Israeli Academy for Fundamental Research for a grant.
R. S. thanks R. Mohapatra and Y. Kamyshkov for related discussions at Snowmass
2001. The research of R. S. was partially supported by the NSF grant
PHY-0098527.

\vfill
\eject


\begin{thebibliography}{99}
\bibitem{c1}{A different class of models features extra dimensions in which SM
fields propagate but without strong localization of fermion wavefunctions; for
these, $\Lambda_L$ can be much lower because of the absence of NFCC processes
\cite{dpq,acd,ta}.  The model of \cite{ta} illustrates how special symmetries
specific to a particular compactification can suppress proton decay and $n-\bar
n$ oscillations adequately for $\Lambda_L \sim O(1)$ TeV.  Our bound applies to
generic models without the assumption of special symmetries.}
\bibitem{lgt}{It would be useful to calculate the $\langle \bar n | {\cal O}_i
| n \rangle$ matrix elements using lattice QCD; to our knowledge, this has
not yet been done.}
\end{thebibliography}
\end{document}